\documentclass[pdflatex,sn-nature]{sn-jnl}

\usepackage{float}
\usepackage{microtype}
\usepackage{graphicx}%
\usepackage{multirow}%
\usepackage{amsmath,amssymb,amsfonts}%
\usepackage{amsthm}%
\usepackage{mathrsfs}%
\usepackage[title]{appendix}%
\usepackage{xcolor}%
\usepackage{textcomp}%
\usepackage{manyfoot}%
\usepackage{booktabs}%
\usepackage{algorithm}%
\usepackage{algorithmicx}%
\usepackage{algpseudocode}%
\usepackage{listings}%
\usepackage{lineno}
\newcommand{\apjl}{Astrophys. J. Lett.}
\newcommand{\icarus}{Icarus}
\newcommand{\mnras}{Mon. Not. R. Astron. Soc.}
\newcommand{\apj}{Astrophys. J.}
\newcommand{\apjs}{Astrophys. J. Suppl. Ser.}
\newcommand{\aj}{Astron. J.}
\newcommand{\planss}{Planet. Space Sci.}

\newcommand{\aap}{Astron. Astrophys.}

\newcommand{\grl}{Geophys. Res. Lett.}
\newcommand{\nat}{Nature}
\newcommand{\psj}{Planet. Sci. J.}

\newcommand{\nar}{New Astron. Rev.}
\newcommand{\ssr}{Space Sci. Rev.}

\newcommand{\jgr}{J. Geophys. Res.}

\theoremstyle{thmstyleone}%
\theoremstyle{thmstyletwo}%

\theoremstyle{thmstylethree}%

\begin{document}

\title[Retention of primordial atmospheres through giant impacts]{\centering Retention of primordial atmospheres through giant impacts \\[0.5em]
{\large\normalfont Accepted in \textit{Nature Astronomy}}}


\author*[1,2]{\fnm{Darius} \sur{Modirrousta-Galian}}\email{modirrousta-galian@sjtu.edu.cn}

\author[3]{\fnm{Jun} \sur{Korenaga}}

\affil[1]{\small \orgdiv{Tsung-Dao Lee Institute}, \orgname{Shanghai Jiao Tong University}, \orgaddress{\street{1 Lisuo Road}, \city{Shanghai}, \postcode{201210}, \country{China}}}

\affil[2]{\small \orgdiv{School of Physics and Astronomy}, \orgname{Shanghai Jiao Tong University}, \orgaddress{\street{800 Dongchuan Road}, \city{Shanghai}, \postcode{200240}, \country{China}}}

\affil[3]{\small \orgdiv{Department of Earth \& Planetary Sciences}, \orgname{Yale University}, \orgaddress{\street{210 Whitney Ave}, \city{New Haven}, \postcode{06511}, \state{CT}, \country{USA}}}

\abstract{Rocky planets the size of Earth or larger are expected to undergo one or more giant impacts during the final stages of their formation. Some of these collisions occur after protoplanetary disk dissipation, potentially driving the loss of primordial hydrogen-rich atmospheres through elevated surface temperatures. However, whether giant impacts promote atmospheric loss or retention remains unclear because they both heat planetary interiors and increase atmospheric gravitational binding. Here we model the evolution of rocky planets after their largest post-disk giant impact, accounting self-consistently for the concurrent increase in gravitational binding energy and mantle thermal energy. Whereas the entire accreted mass contributes to the increase in gravitational binding energy, only part of the impact energy raises mantle temperature. Mantle melting consumes energy as latent heat, while the impactor's metallic core transfers little energy to the mantle and remains superheated at depth. Giant impacts therefore increase gravitational binding energy more than mantle thermal energy, shifting planets towards atmospheric retention. Larger, atmosphere-bearing planets emerge preferentially through collisional growth. By contrast, smaller planets without such impacts tend to lose their primordial atmospheres under sustained stellar irradiation. At the population level, these results suggest that sub-Neptunes have, on average, more massive rocky interiors than super-Earths.}



\maketitle

\section{Introduction}

Observations reveal a bimodal distribution of exoplanet radii, with the two maxima at $1.3~R_{\oplus}$ and $2.4~R_{\oplus}$ and the minimum at $1.75~R_{\oplus}$ \cite{Fulton2017,Fulton2018}. The two maxima have been attributed to rocky planets without (super-Earths) and with (sub-Neptunes) primordial atmospheres, respectively, requiring some exoplanets to have lost large amounts of primordial hydrogen, which plays a crucial role in various physical and chemical processes and strongly influences a planet’s bulk properties and observable traits \cite{Lopez2014}. One popular explanation for the observed bimodality is that giant impacts occurring after the dispersal of the protoplanetary disk remove primordial envelopes \cite{Biersteker2019,Izidoro2022,Chance2022}; such impacts cause mantle shock heating, and the resulting high surface temperatures drive atmospheric escape through a process termed ``core powered mass loss'' \cite{Ginzburg2018,Gupta2019}. The core powered mass loss framework itself does not specify how such high surface temperatures are generated; rather, it describes the ensuing atmospheric escape once they are present. In the post-disk regime considered here, giant impacts provide the relevant source of heating. Post-disk giant-impact models have therefore treated such impacts as atmospheric-removal events that help convert initially atmosphere-bearing planets into super-Earths. Our simulations test this premise by evaluating impact heating together with the concurrent increase in the planet's nuclear mass, which previous calculations do not self-consistently incorporate into the atmospheric binding energy at the same stage of the calculation \cite{Biersteker2019,Chance2022,Izidoro2022}. This coupling is important because accretion does not merely heat the planet: it also adds mass, deepening the gravitational potential well and making the post-impact atmosphere less prone to hydrodynamic escape. Existing models also overlook that a significant fraction of the impact energy is partitioned into processes that do not raise the planet’s surface temperature such as in mantle melting~\cite{Miyazaki2019b} and core segregation~\cite{Dahl2010,Korenaga2023b,Zhou2024}. Any realistic evaluation of impact-driven atmospheric loss must therefore account for all of these effects simultaneously.

In this paper, we show that atmospheric evolution during giant impacts is governed by the increase in planetary gravitational binding energy, which overshadows the heating effects of giant impacts. We first note that the total energy involved in a giant impact event is well approximated by the change in the gravitational binding energy of the system before and after the collision. N-body simulations show that giant impacts typically occur at velocities near the mutual escape velocity ($v_{\rm esc}$) \cite{Agnor1999,Kokubo2007,Chambers2013,Suli2021,Matsumoto2021}, and smoothed particle hydrodynamics simulations show that impacts above ${\sim}1.2~v_{\rm esc}$ increasingly result in hit-and-run events with partial accretion and substantial mantle mass loss \cite{Genda2012,Asphaug2010,Gabriel2020}. In such cases, not all kinetic energy is transferred to the target, with much of it escaping with the impactor remnant. Therefore, even when the impact velocity exceeds the escape velocity, the energy actually deposited into the planet remains close to that associated with a mutual-escape velocity collision \cite{Dou2024}. Mechanical stripping during the impactor’s passage through the atmosphere removes only a small fraction of the gas \cite{Okeefe1982}, becoming progressively less effective for larger impactors because of reduced viscous coupling at higher Reynolds numbers (see online methods). However, the collision also launches a shock through the planetary interior, driving global ground motion and a secondary shock into the atmosphere. This process can remove a substantial fraction of the atmosphere, with shock-driven losses often of order $10\%{-}20\%$ \cite{Walker1986,Genda2003,Schlichting2015,Kegerreis2020,Matsumoto2021,Roche2025}, though larger losses are possible depending on the impact conditions, such as the impactor mass, entry speed, and impact angle \cite{Inamdar2015,Venturini2020}. Whereas successive impacts could in principle increase the cumulative loss, recent formation models suggest that post-disk giant impacts are too infrequent to efficiently remove atmospheres through this channel alone \cite{Ogihara2020}. Consequently, the dominant process affecting the atmospheric evolution is the net change in the planet’s gravitational binding energy rather than the localized mechanical effects of the collision itself. In an idealized scenario where all deposited energy were available for heating, the gain in gravitational binding energy would balance the thermal energy delivered to the mantle, and the two effects would approximately cancel. Simulations instead suggest that the metallic core of the impactor does not contribute to shock heating; upon impact, the core sinks rapidly toward the center of the merged planet, penetrating through the silicate mantle with minimal energy transfer and depositing its energy deep in the interior \cite{Dahl2010,Korenaga2023b,Zhou2024}. As a result, only the kinetic energy of the impactor’s silicate mantle is available to generate shock heating. However, even this process is not fully efficient because a substantial fraction of the impact energy is consumed as latent heat during melting \cite{Miyazaki2019b}, reducing the energy available to raise the mantle’s temperature. It follows that, whereas the gravitational binding energy increases fully with the added mass, only a fraction of the impact energy is available to heat the atmosphere. Consequently, the atmosphere becomes more gravitationally bound following a giant impact.

\textls[-4]{Giant impacts can significantly heat a planet’s surface. However, this energy is largely damped in the upper atmosphere. Below the radiative-convective boundary, the temperature follows an adiabatic profile set by the temperature across the magma ocean boundary layer \cite{Modirrousta2023}. Above it, convection ceases, radiative diffusion dominates, and the temperature profile decouples from the surface. Consequently, surface heating has little influence on the upper layers that control atmospheric escape, which are instead governed by conditions near the sonic point where gas becomes transonic \cite{Hunten1982,Volkov2011a,Volkov2013}. The structure of the escaping flow depends on whether the region below the sonic point is optically thick or thin \cite{Modirrousta2023}. In the optically thick case, the gas behaves similarly to a blackbody and cools efficiently; the atmosphere approaches an isothermal structure at the planetary equilibrium temperature \cite{Hubeny2003,Hansen2008,Guillot2010}, with minor cooling near the sonic point \cite{Modirrousta2025b}. In the optically thin case, high-energy photons heat the rarefied gas, forming a thermosphere with a steep temperature gradient governed by thermal conduction \cite{Bates1951,Bates1959,Gross1972}. In both scenarios, the temperature at the sonic point is determined by stellar irradiation, not by the surface temperature \cite{Tang2024,Modirrousta2025b}. The only exception arises if the magma ocean is hot enough for convection to extend all the way to the sonic point \cite{Modirrousta2023}. In that limit, the magma ocean and the sonic point become thermally coupled, but the temperatures needed to sustain such a configuration are generally unattainable \cite{Modirrousta2025b}. As a result, surface heating from giant impacts has little effect on escape rates. In contrast, accretion increases the planet’s gravitational binding energy, deepening the potential well and reducing atmospheric escape relative to the planet’s lower-mass pre-impact state.}

When gravitational binding energy, core segregation, and the enthalpy of fusion are accounted for, the canonical view of giant impacts as drivers of atmospheric loss is reversed (Figure~\ref{fig:schematic}). Giant impacts deposit most of their energy into the silicate mantle, generating shock heating. However, only a small fraction of this energy is directly transferred to the upper atmosphere and contributes to escape. In contrast, the accompanying increase in gravitational binding energy affects the entire atmosphere and dominates at high altitudes, resulting in a net stabilizing effect. We quantify these competing effects using a revised core powered mass loss model that couples hydrodynamic escape with realistic mantle thermodynamics during giant impacts, and we applied this model to a synthetic population of exoplanets.

\section{Results}
\subsection{Reproducing the bimodal radius distribution}

After evolving a synthetic population of $10^8$ planets in the aftermath of their post-disk giant impact phase, we determine the stability of their primordial atmosphere by comparing three timescales: magma ocean cooling, atmospheric loss, and final planetary age. If a planet undergoes a giant impact, its mass increases by that of the impactor (i.e., assuming complete accretion), and its surface temperature is elevated by the effective deposited energy. We evaluate the atmospheric mass-loss rate during the post-impact magma-ocean phase. After the magma ocean crystallizes, or if it was never impacted, we evaluate atmospheric mass loss under equilibrium-temperature conditions over billion-year timescales (i.e., the final planetary age). The adopted giant-impact size distributions are shown in Extended Data Figure~\ref{fig:chi_distribution}. Our model successfully replicates the overall shape of the observed distribution, with maxima at approximately $1.3~R_{\oplus}$ and $2.4~R_{\oplus}$ and the minimum at $1.75~R_{\oplus}$ for all impact distributions and impact probability combinations (Figure~\ref{fig:bimodal} and Extended Data Figure~\ref{fig:bimodal_2}). All initial parameters, including mass, composition, temperature, and impact histories, are drawn from observationally motivated distributions.

\subsection{Atmospheric retention after giant impacts}
\textls[-3]{Figure~\ref{fig:impact_probability} shows that the likelihood of complete atmospheric loss decreases as both the initial planet mass and the impactor-to-final-planet mass ratio $\chi$ increase. This trend emerges because both greater initial planetary mass and larger impactors increase the planet’s gravitational binding energy, suppressing atmospheric escape. The curve defined by $\chi{=}0.58{-}0.0188\left(M_{\rm ini}/M_{\oplus}\right)^{1.794}$ (red dashed line in Figure~\ref{fig:impact_probability}) represents the analytic approximation for the impactor-to-final-planet mass ratio required for 1\% of all planets of initial mass $M_{\rm ini}$ to experience total atmospheric loss. Planets that experienced giant impacts with $\chi$ larger than this curve almost always retain their primordial atmospheres. As shown in Figure~\ref{fig:bimodal} and Extended Data Figure~\ref{fig:bimodal_2}, our results are robust across all reasonable choices of $\chi$ distribution and post-disk giant impact probability. The adopted initial planetary mass distribution is shown in Figure~\ref{fig:mass_distribution}a. The mantle adiabats and impact-energy partitioning used to calculate post-impact heating are shown in Figure~\ref{fig:mantle}, and the corresponding post-impact surface temperatures are shown in Figure~\ref{fig:T_sur}. Similarly, varying the metallic core mass fraction between 10\% and 40\% has little effect on the outcome. This range is consistent with constraints set by stellar elemental ratios, which suggest that exoplanet core mass fractions typically fall between 20\% and 40\%, with a broader possible range of 5\% to 55\% \cite{Unterborn2019}, and by geophysical modeling of rocky exoplanet densities, which yields an average of 26$\pm$7\% \cite{Zeng2016}. For the initial atmospheric mass fraction, we adopt a normalized standard power-law scaling with nucleus mass \cite{Lee2015,Ginzburg2016,Mordasini2020}, with parameters informed by planetary formation theory and calibrated to match the observed clustering of $7{-}10~M_{\oplus}$ planets in the $2.4~R_{\oplus}$ radius maximum \cite{Muller2024,Parc2024}. The control case without post-disk giant impacts (Extended Data Figure~\ref{fig:bimodal_2}g) shows atmospheric loss under stellar irradiation alone. However, this case overproduces super-Earths and underproduces sub-Neptunes, indicating that too many planets lose their hydrogen-rich atmospheres. The opposite occurs in the control case where every planet experiences a post-disk giant impact (Extended Data Figure~\ref{fig:bimodal_2}h), with too many planets retaining hydrogen-rich atmospheres and entering the $2.4~R_{\oplus}$ radius maximum, and too few becoming atmosphere-poor super-Earths. Together, these limiting cases show that giant impacts regulate the relative populations of the two maxima by shifting planets across the mass threshold for atmospheric retention. This behavior follows from the primary control of planetary mass on atmospheric escape: planets above a critical mass preferentially retain hydrogen-rich envelopes, whereas lower-mass planets tend to lose them under sustained stellar irradiation. This reverses the standard interpretation of giant impacts: Rather than acting as a competing atmospheric-loss mechanism to photoevaporation, they promote atmospheric retention and help build the sub-Neptune population by biasing the planetary population across the mass threshold that separates super-Earths from sub-Neptunes.}

\section{Discussion}

\textls[-3]{Our results indicate that giant impacts help planets sustain, rather than remove, hydrogen-rich atmospheres relative to their lower-mass pre-impact states. This stabilizing effect suggests that sub-Neptunes, which occupy the larger ${\sim}2.4~R_{\oplus}$ radius maximum, likely underwent more extensive collisional growth and developed more massive rocky interiors than super-Earths, which occupy the smaller ${\sim}1.3~R_{\oplus}$ maximum. Statistical analyses of exoplanets with measured masses and radii show that sub-Neptunes have masses above ${\sim}4{-}5~M_{\oplus}$, whereas super-Earths have masses below this threshold \cite{Swain2019,Muller2024,Parc2024}. This transition mass marks the point where planetary nuclei begin to retain hydrogen-rich envelopes, which is consistent with the idea that more massive rocky interiors are better able to accrete and retain hydrogen-rich envelopes~\cite{Ida2004,Ikoma2012,Lee2015}. In this view, the bimodal radius distribution reflects a threshold in atmospheric escape set by planetary mass, with giant impacts acting as a key evolutionary pathway for planets to cross this threshold. Current and planned missions such as JWST, Ariel, and Twinkle will test this framework by using transmission spectroscopy to detect hydrogen-rich atmospheres in planets in the second maximum of the bimodal radius distribution. At the same time, the PLATO and JUST missions will measure planetary masses through the radial velocity method, and the Nancy Grace Roman Space Telescope will use microlensing to extend mass measurements to smaller and more distant planets. Collectively, these missions will test whether sub-Neptunes are indeed systematically more massive than super-Earths and whether they tend to retain hydrogen-rich atmospheres.}

\section{Methods}
\label{sec:methods}

After the dissipation of the protoplanetary disk, planets cannot reaccrete a new hydrogen-rich atmosphere following a giant impact, making any atmospheric loss permanent. We thus focus on modeling the mass loss from the largest post-disk giant impact because this determines the final state of the atmosphere. In the following, we describe our method for modeling the atmosphere, equilibrium temperature, age, mass, mantle, and post-impact conditions of our planet population.

\subsection{Atmospheric model}
\label{sec:atm_model}

Following a giant impact, the planetary nucleus is hot and largely molten, with the atmosphere above having a temperature profile that is inherently unstable and hydrodynamic. Because of mass conservation, the mass flux can be determined at any radial distance. For simplicity, it is usually evaluated at the sonic point, where advection velocity equals the sound speed $c_{\rm s}$. The mass loss rate is,
\begin{equation}
    \frac{{\rm d}M_{\rm atm}}{{\rm d}t} = 4 \pi R_{\rm s}^{2} \rho_{\rm s} c_{\rm s},
\label{eq:mass_cons}
\end{equation}
with $R_{\rm s}$ and $\rho_{\rm s}$ being the radius and density at the sonic point. The sonic point is given by,
\begin{equation}
    R_{\rm s}=\frac{GM_{\rm n}\bar{\mu}}{2k_{\rm B}T_{\rm s}},
\end{equation}
where $G$ is the gravitational constant, $M_{\rm n}$ is the mass of the planetary nucleus (which we approximate as the total planetary mass), $\bar{\mu}$ is the mean molecular weight (Equation~\ref{eq:mu}), $k_{\rm B}$ is the Boltzmann constant, and $T_{\rm s}$ is the temperature at the sonic point. If the post-impact surface temperature is sufficiently high, convection extends from the planetary surface to the sonic point, and $T_{\rm s}$ is then set by the surface adiabat. In contrast, if the adiabatic temperature profile falls below the equilibrium-temperature profile before reaching the sonic point, a radiative layer develops below the sonic point and $T_{\rm s}$ is instead set by a mixture of radiative equilibrium and adiabatic cooling from advection, which is approximately $T_{\rm s} {\approx} 2T_{\rm eq}/(\gamma{+}1)$ \cite{Modirrousta2025b}, with $T_{\rm eq}$ being the planetary equilibrium temperature and $\gamma$ the heat capacity ratio of the gas mixture (see discussion on Equation~\ref{eq:mu}). In practice, we evaluate both limiting temperatures and adopt the larger of the two, which selects the appropriate regime. Solving the Euler equations for planetary outflow shows that the atmospheric escape rate is determined by Equation~\ref{eq:mass_cons} rather than by the ``energy limited'' argument, which posits that mass loss is constrained by the luminosity escaping through the radiative-convective boundary (see Equation~2 in Ref.\,\cite{Modirrousta2025b}). Equation~\ref{eq:mass_cons} does not require the outflow to remain thermally coupled to the cooling luminosity of the interior because the gas can instead expend its own internal energy and expand adiabatically as it passes through the sonic region. Independent calculations likewise support this interpretation \cite{Tang2024}. However, evaluating Equation~\ref{eq:mass_cons} for a large exoplanet population is computationally demanding because the sonic point density depends on the full atmospheric structure and cannot be determined analytically. Therefore, we instead adopt the following analytic fit \cite{Modirrousta2025b},
\begin{equation}
    \frac{{\rm d}M_{\rm atm}}{{\rm d}t}=1.26{\times}10^{15}\left(\frac{R_{\rm s}}{R_{\oplus}}\right)^{2}\exp\left({-}\frac{19}{27}\frac{R_{\rm s}}{R_{\oplus}}\right)\quad {\rm kg~s^{-1}},
\label{eq:fit}
\end{equation}
which reproduces simulation results with a coefficient of determination of $R^{2}{=}0.93$. Equation~\ref{eq:fit} is based on the boundary-layer analysis of Ref.\,\cite{Modirrousta2025b}, which treats heat transport through the atmospheric boundary layer as conductive. This approximation is adequate for the high post-impact surface temperatures considered in this study. We note, however, that for cooler atmospheres, radiative transport may become important in the boundary layer, requiring a revised fit that includes both conductive and radiative heat transport. We do not include X-ray and ultraviolet induced photoevaporation in our framework because this mechanism requires a thermosphere \cite{Bates1951,Bates1959,Gross1972}, which forms only after the planet cools sufficiently for the upper atmosphere to become optically thin to high-energy photons \cite{Modirrousta2023}. In the aftermath of a giant impact, however, surface temperatures are high and the sonic point lies deep within the atmosphere, below the photosphere where the gas becomes optically thick. Under these conditions, the sonic point is optically thick, and any heating from X-ray and ultraviolet photons is offset by efficient blackbody cooling \cite{Modirrousta2023}. For planets that do not experience a post-disk giant impact, internal temperatures would be sufficiently low for a thermosphere to form, enabling X-ray- and ultraviolet-induced photoevaporation. Incorporating this mechanism would further increase the outflow rate of non-impacted planets, reinforcing our conclusion that planets experiencing giant impacts are more likely to retain their atmospheres.

Our model uses three timescales. The first is the mass loss timescale $t_{\rm loss}{=}M_{\rm atm}/\dot{M}_{\rm atm}$, where we assume a constant mass loss rate $\dot{M}_{\rm atm}$ calculated from Equation~\ref{eq:fit} at the moment of impact. The assumption of a constant mass loss rate allows us to estimate the characteristic timescale of atmospheric loss, even though the actual mass loss rate would evolve over time.

The second is the magma ocean cooling timescale, which provides the characteristic time during which a magma ocean remains sufficiently luminous to drive atmospheric loss. A magma ocean can cool efficiently until its surface reaches the rheological transition, at which point the melt fraction decreases to $\phi{\approx}0.4$ and convection becomes sluggish \cite{Solomatov2015}. The cooling timescale depends on the initial temperature of the magma ocean and the thermal blanketing effect of the atmosphere \cite{Hayashi1979,Hubbard1980,Stevenson1982}. Based on a thermochemically self-consistent simulation that includes crystal accumulation and matrix compaction, Earth’s magma ocean is estimated to have cooled in approximately ${\sim}10~{\rm kyrs}$ \cite{Miyazaki2019b}. This represents a lower bound because the simulation does not include thermal blanketing from the atmosphere. To generalize this timescale to larger planets, we scale it using the post-impact energy increase given by Equation~\ref{eq:dE}. We divide this energy by the planetary surface area, $R_{\rm n}^2$, to estimate the corresponding cooling-rate scaling, and normalize the result to the estimated cooling time of Earth’s primordial magma ocean, yielding $t_{\rm cool} {=} 66.5 \left(M_{\rm n}/M_{\oplus}\right)^{5/4}[1{-}\left(1{-}\chi\right)^{7/4}]{\rm F}_{2}\left(f_{\rm c}\right)~{\rm kyr}$. Here, $\chi$ is the ratio of the impactor mass to the final planetary mass, and $F_{2}(f_{\rm c})$ is a dimensionless function of the metallic core mass fraction $f_{\rm c}$ that accounts for the internal partitioning of gravitational binding energy in a differentiated planet. The parameters $\chi$, $f_{\rm c}$, and $F_{2}$ are defined in the post-impact conditions section below (Equations~\ref{eq:dE} and \ref{eq:F_2}), where we derive the deposited energy from giant impacts in detail. This timescale is intended to describe the cooling of a rocky nucleus with a global magma ocean. It is therefore an approximation when applied to planets with substantial hydrogen envelopes, which can affect the radius, luminosity, and thermal blanketing of the planet. A fully self-consistent treatment of such planets would require a coupled interior-envelope cooling model. In Extended Data Figure~\ref{fig:bimodal_2}i, we show that lengthening the magma ocean cooling time, equivalent to imposing atmospheric thermal blanketing, has little effect on the final shape of the bimodal distribution because, as explained earlier, high internal magma temperatures are damped in the upper atmosphere.

The third timescale is the final planetary age. Data on the ages of planet-hosting stars is limited \cite{Christensen2018}: an analysis of 33 Kepler planet-candidate host stars found that most have ages between 3 and 6~Gyr, with large uncertainties \cite{Aguirre2015}. Additional studies indicate that such stars are rarely younger than 0.5~Gyr or older than 11~Gyr \cite{Bonfanti2015}, and have a mean age of approximately 5~Gyr \cite{Bonfanti2016}. Assuming that stellar age is coeval with that of their planets, we set an average planetary age of $5~{\rm Gyr}$.

\textls[-7]{The fate of a planet’s atmosphere is determined as follows. We first assess whether the planet undergoes a post-disk giant impact. If not, we calculate a cold atmospheric mass loss timescale $t_{\rm loss,cold}$ using Equation~\ref{eq:fit} with surface temperature set at the planet’s equilibrium temperature, and compare it to the final planetary age $t_{\rm age}{=}5~{\rm Gyr}$. If $t_{\rm age}{>}t_{\rm loss,cold}$, the atmosphere is removed and the transit radius corresponds to the planetary nucleus radius (Equation~\ref{eq:R_p}); otherwise, the atmosphere is partially retained and updated by subtracting the expected loss over 5 Gyr, with the final radius then recalculated from the evolved atmospheric mass. If the planet does undergo a giant impact, its mass is increased assuming complete accretion of the impactor. We compare the hot atmospheric mass-loss timescale $t_{\rm loss,hot}$ to the magma-ocean lifetime $t_{\rm cool}$. If $t_{\rm cool}{>}t_{\rm loss,hot}$, the atmosphere is removed during the magma-ocean phase and the transit radius is set to the planetary nucleus radius (Equation~\ref{eq:R_p}). Otherwise, the atmosphere is reduced by the expected loss over $t_{\rm cool}$. The magma ocean then cools to the rheological transition, the hot phase ends, and the subsequent evolution is followed using the cold mass-loss timescale over the remaining time interval, $\Delta t{=}t_{\rm age}{-}t_{\rm cool}$. If $\Delta t{>}t_{\rm loss,cold}$, the remaining atmosphere is removed and the transit radius is set equal to that of the planetary nucleus (Equation~\ref{eq:R_p}). If $\Delta t{<}t_{\rm loss,cold}$, the atmosphere is partially retained and updated accordingly, and the final radius is recalculated from the evolved atmospheric mass. This treatment is a simplified two-stage approximation to a continuously evolving escape history. Resolving the full time dependence would require a self-consistent thermal-evolution calculation for the interior and atmosphere of each planet in the synthetic population, which is computationally prohibitive for a population study.}

For planets that retain hydrogen-rich atmospheres, the transit radius is more difficult to determine because of its sensitivity to the planet’s thermodynamic state \cite{Lopez2014} and the transparency of atmospheric gases \cite{Fortney2005,Betremieux2017,Heng2017}, the latter of which varies strongly with both temperature and composition \cite{Lenzuni1991,Freedman2008,Freedman2014}. To avoid these complexities, we use Equations 3 and 4 of Ref.\,\cite{Lopez2014}, which are analytic approximations for the transit radius of hydrogen-rich primordial atmospheres based on global planetary models with non-gray opacities. We modify this parameterization to account for geological outgassing, which alters both the composition and structure of the atmosphere. The hydrogen mass fraction,
\begin{equation}
X_{\rm H} = \frac{f_{\rm a}}{f_{\rm a} + 2{\times}10^{-4}(1 - f_{\rm c})},
\end{equation}
is computed using the accreted hydrogen mass fraction $f_{\rm a}$, the core mass fraction $f_{\rm c}$, and a constant $2{\times}10^{-4}$ that represents the mass of outgassed volatiles per unit mantle mass \cite{Kasting1986,Liu2004,Hirschmann2009,Marty2012}. We estimate the atmospheric mass fraction using,
\begin{equation}
    f_{\rm a} = \frac{5.5{\times}10^{-5}\left(\frac{M_{\rm n}}{M_{\oplus}}\right)^{2}}{1+5.5{\times}10^{-5}\left(\frac{M_{\rm n}}{M_{\oplus}}\right)^{2}},
\label{eq:scaling_atm}
\end{equation}
which follows a power-law scaling with nucleus mass, as suggested in previous studies \cite{Lee2015,Ginzburg2016,Mordasini2020}, and asymptotically approaches unity for large $M_{\rm n}$. We adopt a parsimonious but physically motivated scaling for the atmospheric mass fraction. The quadratic dependence on nucleus mass is based on the Bondi accretion rate, which scales with the square of the planetary mass and determines the rate of gas capture from the protoplanetary disk \cite{Bondi1952,Edgar2004}. The prefactor $5.5{\times}10^{-5}$ is chosen to reproduce the observed clustering of $7{-}10~M_{\oplus}$ planets around the $2.4~R_{\oplus}$ radius maximum \cite{Muller2024,Parc2024}. Differences in the numerical constants between our study and previous studies arise from the adopted equations of state. Equation~\ref{eq:scaling_atm} and all other fits presented in this paper only serve as representative scalings for exoplanet properties. 

The metallic core mass fraction is randomly sampled from a uniform distribution of 10\% to 40\%. This range is consistent with the constraints set by stellar Fe/(MgO+SiO$_{2}$) ratios \cite{Unterborn2019} and rocky exoplanet bulk densities \cite{Zeng2016}. We assume a mean molecular weight of $\bar{\mu}_{\rm H}{=}2.3~{\rm amu}$ for the hydrogen-rich component and $\bar{\mu}_{\rm z}{=}31~{\rm amu}$ for the outgassed component (average of carbon dioxide and water vapor). The total mean molecular weight of the atmosphere is then given by,
\begin{equation}
\bar{\mu} = \left( \frac{X_{\rm H}}{\bar{\mu}_{\rm H}} + \frac{1 - X_{\rm H}}{\bar{\mu}_{\rm z}} \right)^{-1}.
\label{eq:mu}
\end{equation}
Using the parameters defined above, we introduce the dimensionless scaling parameter,
\begin{equation}
S = X_{\rm H} + (1 - X_{\rm H})\frac{\bar{\mu}_{\rm H}}{\bar{\mu}_{\rm z}},
\end{equation}
which quantifies how the relative enrichment in heavier outgassed species reduces the thickness of the inner convective region relative to a primordial atmosphere. A higher contribution from outgassed heavy species increases the mean molecular weight of the atmosphere and reduces its radial extent relative to a purely primordial H/He envelope. Whereas the initial inventory of volatiles stored in mantle-forming rocks is thought to be well constrained by geochemical analyses \cite{Kasting1986,Liu2004,Hirschmann2009,Marty2012}, the amount that actually outgasses into the atmosphere is highly sensitive to the convective mode of the mantle. Venus, for example, has a large secondary atmosphere because it experiences stagnant lid convection \cite{Solomatov1996}, thus being unable to reintroduce these volatiles back into the mantle after being outgassed. In contrast, Earth has plate tectonics \cite{Korenaga2013}, and it is therefore able to form carbon \cite{Isson2020} and water \cite{Korenaga2017} cycles that recycle these volatiles across the mantle and atmosphere continuously. If the primordial atmosphere is too thick and the magma cannot crystallize to form a mantle, the magma may not outgas fully and instead reach an equilibrium with the atmosphere set by Henry’s law \cite{Modirrousta2025a}. Moreover, hydrodynamic escape itself is fractionating, leading to a relative enrichment in the atmospheric composition \cite{Hunten1987,Pepin1997,Modirrousta2024}. Variations in the atmospheric metallicity would modify both the transit radius and the susceptibility to atmospheric escape, which may be observable through atmospheric spectroscopy \cite{Modirrousta2024}. Such detailed analyses, though important, fall outside the remit of this study. To keep the framework tractable, we assume that planets outgas their volatile inventory efficiently upon crystallization and that subsequent hydrogen loss does not substantially fractionate the retained heavy species. This simplification is consistent with geodynamical models that suggest that stagnant lid convection is the most likely mode of mantle convection \cite{Solomatov1997,Korenaga2010}, atmospheric loss models that suggest that heavy species are resilient to atmospheric loss \cite{Modirrousta2024}, and with JWST observations that suggest that some sub-Neptunes have high metallicities \cite{Piaulet2024,Coulombe2025}. The total transit radius is then given by $R_{\rm p}{=}R_{\rm n}{+}\Delta R_{\rm conv}{+}\Delta R_{\rm rad}$, where $R_{\rm n}$ is given by Equation~\ref{eq:R_p},
\begin{equation}
    \Delta R_{\rm conv} = 2.06SR_{\oplus}\left(\frac{M_{\rm n}}{M_\oplus}\right)^{-0.21}\left(\frac{f_a}{0.05}\right)^{0.59} \left(\frac{T_{\rm eq}}{255}\right)^{0.176}\left(\frac{5}{\rm Age}\right)^{-0.18},
\label{eq:R_conv}
\end{equation}
and,
\begin{equation}
    \Delta R_{\rm rad} = \frac{9k_{\rm B} T_{\rm eq}}{\bar{\mu} g},
\end{equation}
with $g$ being the local gravitational acceleration and the age being in Gyr \cite{Lopez2014}.

\subsection{Planetary equilibrium temperature distribution}
\label{sec:app_temp}

We sample the equilibrium temperature distribution of exoplanets using Table~4 of Ref.\,\cite{Fulton2018}, which lists planetary radii and statistical weights for systems meeting the same selection criteria as Ref.\,\cite{Fulton2017}. Each planet is identified from the full stellar and planetary dataset in their Table~3, and the statistical weights are applied to compute the weighted mean and standard deviation. The resulting distribution of equilibrium temperatures (assuming zero albedo) is well approximated by a Gaussian with a mean of 1311~K and a standard deviation of 441~K. To avoid extrapolating beyond the dataset, we impose lower and upper temperature bounds of 495~K and 2060~K, respectively.

\subsection{Planetary initial mass distribution}

Accurately determining the mass distribution of exoplanets is challenging because of the limited number of mass measurements available. In this study, we do not aim to infer the intrinsic properties of the bimodal radius distribution, but rather to demonstrate that giant impacts help planets retain their primordial hydrogen-rich atmosphere relative to their lower-mass pre-impact state. We therefore adopt a standard mass distribution and assume a broken power-law form \cite{Ginzburg2018}, without attempting to refine it,
\begin{equation}
    P(M_{\rm n}) = 
    \begin{cases}
    0.0916,& \text{if } 1 {\leq} \frac{M_{\rm n}}{M_{\oplus}}{<} 6.4\\
    0.0918\left[\left(\frac{M_{\rm n}}{6.4~M_{\oplus}}\right)^{-2}-\left(\frac{87 M_{\oplus}}{6.4~M_{\oplus}}\right)^{-2}\right],& \text{if } 6.4 {\leq} \frac{M_{\rm n}}{M_{\oplus}}{<} 87\\
    0,              & \text{otherwise}
    \end{cases}
    ,
\label{eq:mass_distribution}
\end{equation}
which is shown in Figure~\ref{fig:mass_distribution}a.

\subsection{Giant impact size and probability}
\label{sec:impact_size}

We use two kinds of probability to describe our giant impact statistics: (1) The probability of a post-disk giant impact occurring, and (2) the probability distribution function for the mass of the giant impactor given that a giant impact occurs. For the first probability, we adopt a lower bound of 52\%, an upper bound of 74\%, and a midpoint of 63\%, based on N-body simulations of late-stage planetary assembly and dynamical instability following protoplanetary disk dispersal \cite{Chance2022,Izidoro2022}. For the second probability, the size of each giant impact is described with the parameter $\chi$ (the ratio of the largest post-disk impactor’s mass over the final planetary mass). Because we define the impactor as the smaller of the colliding bodies, $0{<}\chi{\leq}0.5$. We expect that the probability density function of parameter $\chi$, $P(\chi)$, decreases monotonically because smaller impactors (small $\chi$) are more common than larger ones (large $\chi$). This also implies $P(\chi{\rightarrow}0){\rightarrow}\infty$. Based on these considerations, we propose the following three monotonically decreasing probability density functions for $\chi$.
The first function is
\begin{equation}
    P_{1}(\chi) = 
    \begin{cases}
    \frac{1}{1.387}\left(\frac{0.5}{\chi}\right)^{1.625}, & \text{if } 0.1 {\leq} \chi{\leq} 0.5\\
    0,              & \text{otherwise}
    \end{cases}
    .
\label{eq:P_chi}
\end{equation}
This function is based on observations of main belt asteroids and near-Earth objects \cite{Stokes2003,Bottke2005} as well as theory \cite{Tanaka1996,OBrien2003}, suggesting that a population of similar bodies in a collisional cascade has a steady-state radius distribution following ${\rm d}n/{\rm d}r{\propto}r^{-3.5}$. If $r{\propto}M^{1/4}$, ${\rm d}n/{\rm d}M{\propto}M^{-1.625}$, whereas if $r{\propto}M^{1/3}$, ${\rm d}n/{\rm d}M{\propto}M^{-1.833}$ \cite{Dohnanyi1969}. The ${\rm d}n/{\rm d}r{\propto}r^{-3.5}$ trend applies only to objects with radii $R{\lesssim}130~{\rm km}$, suggesting that larger bodies have not reached dynamic equilibrium yet \cite{Lissauer2013}. It is unclear if such dynamic equilibrium can be reached for the bodies responsible for giant impacts because planet formation is believed to end in ${\sim}100~{\rm Myr}$ \cite{Quintana2016}, which is short on astronomical timescales. The upper limit of $\chi{=}0.5$ is set because $\chi$ cannot exceed 0.5. The lower limit of $\chi{=}0.1$ is adopted here because this function is not integrable at $\chi{=}0$.

The second function is
\begin{equation}
    P_{2}(\chi) = 
    \begin{cases}
    \frac{40}{11}\left(\frac{0.5}{\chi}\right)^{\frac{1}{11}}\left[1+\left(\frac{0.5}{\chi}\right)^{-\frac{10}{11}}\right]^{-2}, & \text{if } 0 {\leq} \chi{\leq} 0.5\\
    0,              & \text{otherwise}
    \end{cases}
    ,
\label{eq:P_chi2}
\end{equation}
where $\chi {\in} \left[0, \infty \right)$. This is a Pareto type III distribution with a cut-off at $\chi{=}0.5$. The Pareto distribution describes many phenomena across various fields, including physics and earth and planetary sciences \cite{Newman2005}. Whereas the Pareto distribution is integrable for $\chi{\in}\mathbb{R}_{{\geq}0}$, we impose a cut-off at $\chi{=}0.5$.

The third function is
\begin{equation}
    P_{3}(\chi) = 
    \begin{cases}
    2\ln\left(\frac{0.5}{\chi}\right), & \text{if } 0 {\leq} \chi{\leq} 0.5\\
    0,              & \text{otherwise}
    \end{cases}
    .
\label{eq:P_chi3}
\end{equation}
This function satisfies $P(\chi{=}0){=}\infty$ and $P(\chi{=}0.5){=}0$ without needing to explicitly set the value to zero at $\chi{=}0.5$.

The abovementioned distributions, shown in Extended Data Figure~\ref{fig:chi_distribution}, serve only as representative test models because the actual post-disk size distribution for the largest giant impact remains uncertain; we therefore also include a uniform distribution in $\chi$ as a control, which we explore in Extended Data Figure~\ref{fig:bimodal_2}. N-body simulations provide limited clarity on the post-disk giant impact size distribution, but they characterize these events as collisions involving impactors with at least 10\% of the target planet's mass \cite{Izidoro2022}, often reaching comparable masses to the target planet itself \cite{Chance2022}. We focus on the largest giant impact after disk dispersal because post-disk impacts determine the final atmospheric state. Before disk dispersal, a planet remains embedded in the protoplanetary disk and can exchange both mass and energy with the surrounding gas. Atmospheric loss caused by disk-embedded impacts is therefore not necessarily final, because the planet can reaccrete protoplanetary disk gas and its envelope can thermally readjust through recycling flows. The thermal outcome of this readjustment is regime dependent: recycling can suppress cooling for temperate envelopes, but for much hotter post-impact envelopes it can contribute to cooling by replacing hot envelope gas with relatively cooler disk gas \cite{Ormel2015b,Moldenhauer2021,Zhu2021,Modirrousta2025a}. We therefore do not model the cumulative thermal history of all disk-embedded impacts. Instead, we treat the pre-impact state before the final post-disk giant impact as an idealized lower-bound state, as discussed below. After disk dispersal, reaccretion from the protoplanetary disk and gas recycling cease, so atmospheric loss following a giant impact is permanent. The largest post-disk impact is therefore the event most relevant to the final atmospheric stability in our model.

\subsection{Mantle phase diagram}
\label{sec:mantle_phase}

Mantle melting has to be modeled to determine the post-impact conditions of a planet. Unlike simpler endmember materials such as water ice, mantle melting occurs over a wide temperature range that starts at the solidus and ends at the liquidus. This temperature range exists because planetary mantles are composed of various solid solutions, mostly silicates and metal oxides. It is important to incorporate a realistic description of mantle melting for two reasons. First, the internal heat flux practically vanishes when a planet has a surface temperature below the rheological transition (${\sim}$1610~K) because a magma ocean ceases to exist \cite{Solomatov2015,Korenaga2023}. Therefore, even if internal temperatures at great depths are high (${>}$4000~K), atmospheric evaporation arising from internal energy stops. Second, the latent heat of fusion is important because it acts as a heat sink. These two effects are not accounted for in simpler modeling approaches such as treating a planet as an isothermal sphere \cite{Ginzburg2018}. 

We describe our approach for modeling mantle melting in the following. Based on experimental data and theory \cite{Zhang1994,Fiquet2010,Miyazaki2019b}, and assuming that rocky exoplanets have a composition similar to that of Earth, the mantle phase diagram is parameterized as follows ($R^{2}{>}0.99$ goodness of fit),
\begin{equation}
    T_{\rm l}=3351.83+44.034P^{0.795}-\frac{1377.118}{\left[1+\left(\frac{P}{99.055}\right)^{3.315}\right]^{10.851}},
\label{eq:T_l}
\end{equation}
\begin{equation}
    T_{\rm 40}=2554.75+129.307P^{0.633}-\frac{999.16}{\left[1+\left(\frac{P}{32.089}\right)^{32.089}\right]^{0.0524}},
\label{eq:T_40}
\end{equation}
\begin{equation}
    T_{\rm s}=632.615+\frac{643.56}{\left[1+\left(\frac{P}{1.570}\right)^{1.796}\right]^{-0.211}},
\label{eq:T_s}
\end{equation}
where $T_{\rm l}$, $T_{40}$, and $T_{\rm s}$ are the liquidus, the rheological transition temperature, and the solidus temperature in K at pressure $P$ in GPa, respectively. Regarding the melt fraction, $\phi$, we assume that it scales linearly between the solidus ($\phi{=}0$) and the rheological transition temperature ($\phi{=}0.4$ \cite{Solomatov2015}), and the rheological transition temperature and the liquidus ($\phi{=}1$), respectively.

\subsection{Internal mantle pressures}
\label{sec:mantle_pressures}

The degree of partial melting depends on the internal pressure distribution, which we model using a polytropic analysis. We adopt the model of a two layered differentiated (i.e., silicate mantle with an iron core) planet with a constant density in each layer to find the pressure distribution \cite{Sohl2015},
\begin{equation}
    P(r)=\frac{3GM_{\oplus}^{2}}{8\pi R^{4}_{\oplus}}\left(\frac{M_{\rm n}}{M_{\oplus}}\right){\rm F}_{1}\left(r,f_{\rm c}\right),
\label{eq:P_r}
\end{equation}
where
\begin{equation}
\begin{split}
    {\rm F}_{1}(r,f_{\rm c})=\left[1+0.3\left(0.325-f_{\rm c}\right)\right]^{-4}\left\{2f_{\rm c}\left(1-f_{\rm c}\right)\frac{\frac{R_{\rm n}}{r}-1}{1-\left(\frac{R_{\rm c}}{R_{\rm n}}\right)^{3}} \right. & \left. -2\left(1-f_{\rm c}\right)^{2}\left(\frac{R_{\rm c}}{R_{\rm n}}\right)^{3}\frac{\frac{R_{\rm n}}{r}-1}{\left[1-\left(\frac{R_{\rm c}}{R_{\rm n}}\right)^{3}\right]^{2}} \right. \\
    & \left. +\left(1-f_{\rm c}\right)^{2}\frac{1-\left(\frac{r}{R_{\rm n}}\right)^{2}}{\left[1-\left(\frac{R_{\rm c}}{R_{\rm n}}\right)^{3}\right]^{2}} \right\},
\end{split}
,
\label{eq:F_1}
\end{equation}
and $M_{\rm n}$, $M_{\oplus}$, $G$, $r$, $f_{\rm c}$ are the nuclear mass, Earth’s mass, gravitational constant, radial distance, and metallic core mass fraction, respectively. Equations~\ref{eq:P_r} and \ref{eq:F_1} are obtained from Equation~8 of Ref.~\cite{Sohl2015} after rewriting the layer masses and densities in terms of the total nuclear mass and radius $M_{\rm n}$ and $R_{\rm n}$, metallic core mass fraction $f_{\rm c}$, and metallic core radius $R_{\rm c}$. Both equations assume that the radius, $R_{\rm n}$, of a two layered planet scales as,
\begin{equation}
    \frac{R_{\rm n}}{R_{\oplus}}{=}\left[1+0.3\left(0.325-f_{\rm c}\right)\right]\left(\frac{M_{\rm n}}{M_{\oplus}}\right)^{\frac{1}{4}},
\label{eq:R_p}
\end{equation}
which is a fit to geophysical models \cite{Zeng2016}. Note that whereas Equation~\ref{eq:R_p} assumes an isothermal cold interior, the volume of rock changes only weakly with temperature. Evaluating Equation~\ref{eq:alpha} gives volumetric expansions of only ${\sim}10\%$ and ${\sim}0.001\%$ for a temperature increase from 1610 to 10,000~K at fixed pressures of 100 and 1000~GPa, respectively. Thermal expansion therefore does not materially affect the mass-radius scaling adopted here. Equation~\ref{eq:F_1} is a function of the ratio of the metallic core radius to the nuclear radius. The radius of the metallic core is found through a polytropic analysis, where the density of a polytrope scales as $P{\propto}\rho^{n}$, with $n$ being the polytropic index. The internal pressure of the metallic core comes from its mass and the mass of the mantle above, so the equation for the polytrope is changed to $P_{\rm c}{+}\Delta P_{\rm m}{\propto} \rho_{\rm c}^{n}$, where $\Delta P_{\rm m}$ is the pressure contribution from the mantle. Inserting $P_{\rm c}{\propto}M_{\rm c}^{2}/R_{\rm c}^{4}$, $\Delta P_{\rm m}{\propto}M_{\rm m}M_{\rm c}/R_{\rm c}^{4}$, and $\rho_{\rm c}{\propto}M_{\rm c}/R_{\rm c}^{3}$, and solving for the parameter values of Earth and those of an idealized pure iron planet \cite{Zeng2016},
\begin{equation}
    \frac{R_{\rm c}}{R_{\rm n}}{=}\frac{0.8f_{\rm c}^{\frac{1}{4}}}{1+0.3\left(0.325-f_{\rm c}\right)}\left[1+0.6\left(\frac{1-f_{\rm c}}{f_{\rm c}}\right) \right]^{-\frac{1}{8}}.
\label{eq:R_c_R_p}
\end{equation}

\subsection{Mantle adiabat}

The equation for the adiabat, with the effect of mantle melting, is
\begin{equation}
    \left(\frac{\partial T}{\partial P}\right)_{S} = \left(\frac{\partial T}{\partial P}\right)^{0}_{S}+\frac{\Delta H_{\rm f}\left(T,P\right)}{c_{P}\left(T,P\right)}\left(\frac{\partial \phi}{\partial P}\right)_{S},
\label{eq:adiabat_T}
\end{equation}
where $\Delta H_{\rm f}$ is the enthalpy of fusion, $\phi$ is the melt fraction,
\begin{equation}
    \left(\frac{\partial T}{\partial P}\right)^{0}_{S}=\frac{\alpha\left(T,P\right)T}{\bar{\rho}_{\rm m}c_{P}\left(T,P\right)},
\label{eq:adiabat_0}
\end{equation}
the parameter $\bar{\rho}_{\rm m}$ is the average mantle density, and the volumetric thermal expansion coefficient and specific heat are,
\begin{equation}
    \alpha\left(T,P\right)=3.622{\times}10^{-5} \exp{\left(-2.377{\times}10^{-5}T-0.0106P\right)}\quad {\rm K^{-1}},
\label{eq:alpha}
\end{equation}
and
\begin{equation}
    c_{P}\left(T,P\right)=627+0.411T-0.211P\quad {\rm J~K^{-1}~kg^{-1}},
\label{eq:c_p}
\end{equation}
respectively, with the temperature $T$ in K and pressure $P$ in GPa. Equations~\ref{eq:alpha} and \ref{eq:c_p} are based on the thermodynamic calculations of Ref.\,\cite{Miyazaki2019b} as parameterized by Ref.\,\cite{Korenaga2023}. Based on ab initio calculations, the enthalpy of fusion is estimated as ($R^{2}{>}0.99$ goodness of fit for the four conditions listed below \cite{Stixrude2005}),
\begin{equation}
    \Delta H_{\rm f}\left(T,P\right)=8.92{\times}10^{6}-\frac{8.92{\times}10^{6}}{1+\left(\frac{P}{15}\right)^{1.5}}+\frac{6{\times}10^{5}}{1+\left(\frac{T}{5000}\right)^{6}}\quad {\rm J~kg^{-1}}.
\label{eq:enthalpy}
\end{equation}
Equation~\ref{eq:enthalpy} satisfies the following conditions: (1) $\Delta H_{\rm f}(T,P){=}6{\times}10^{5}$~J~kg$^{-1}$ at $P{=}0$~GPa and $T{=}0$~K, (2) $\Delta H_{\rm f}(T,P){=}9{\times}10^{6}$~J~kg$^{-1}$ at $P{=}136$~GPa and $T{=}4500$~K \cite{Miyazaki2019b}, (3) $\Delta H_{\rm f}(T,P){=}0$~J~kg$^{-1}$ at $P{=}0$ GPa and $T{\rightarrow}\infty$, and (4) $\Delta H_{\rm f}(T,P)$ converges to a maximum value as $P{\rightarrow}\infty$ \cite{Stixrude2005}. Figure~\ref{fig:mantle}a shows the adiabat for a $10~M_{\oplus}$ planet with a metallic core mass fraction of 0.325 at various mantle temperatures.

\subsection{Post-impact conditions}
\label{sec:post_impact_conditions}

The effective energy deposited after a giant impact is equal to the gravitational potential energy of the mantle after the impact minus the energy before (assuming perfect accretion and impact velocities at escape velocity \cite{Agnor1999,Kokubo2007,Asphaug2010,Genda2012,Chambers2013,Gabriel2020,Suli2021}). The energy of the metallic core is not included because most of the impactor's metallic core will sink quickly to the center of the planet \cite{Dahl2010,Korenaga2023b,Zhou2024} and remain superheated with inefficient cooling \cite{Miyazaki2019b}. We also do not account for the direct mechanical heating of the atmosphere during impact. This contribution is small, and it can be shown using the Navier-Stokes equations. Assuming incompressibility, the momentum equation reads,
\begin{equation}
    \rho \frac{{\rm D}\vec{v}}{{\rm D}t} = -\nabla P +\rho \vec{g} + \eta \nabla^{2} \vec{v},
\end{equation}
where $v$ is velocity, $t$ is time, $P$ is pressure, $\rho$ is density, $g$ is gravitational acceleration, and $\eta$ is viscosity. The first term is acceleration, the second is pressure gradient, the third is gravity, and the fourth is momentum diffusion. A large fourth term indicates strong internal friction and an efficient conversion of kinetic energy from the impactor to heat in the surrounding fluid. By non-dimensionalizing, we obtain,
\begin{equation}
    \frac{{\rm D}\vec{v}^{\ast}}{{\rm D}t^{\ast}} = -\nabla^{\ast} P^{\ast} +\frac{1}{\rm Fr^{2}} \vec{g} + \frac{1}{\rm Re}\nabla^{\ast 2} \vec{v}^{\ast},
\end{equation}
where $\rm Fr$ is the Froude number and $\rm Re$ is the Reynolds number. The Reynolds number is defined as ${\rm Re}{=}\rho_{\rm a}v_{\rm i}L_{\rm i}/\eta_{\rm a}$, with $\rho_{\rm a}$ and $\eta_{\rm a}$ being the density and viscosity of the atmosphere, and $v_{\rm i}$ and $L_{\rm i}$ being the velocity and characteristic length of the impacting body respectively. Selecting appropriate values of $\rho_{\rm a}{\sim}1~{\rm kg~m^{-3}}$, $\eta_{\rm a}{\sim}10^{-4}~{\rm Pa~s}$, $L_{\rm i}{\sim}10^{6}~{\rm m}$, and $v_{\rm i}{\sim}10^{4}~{\rm m~s^{-1}}$, yields a Reynolds number of ${\sim}10^{14}$, causing the rightmost term to vanish and the fluid to behave inviscidly. In other words, viscosity can be ignored in the high Reynolds number regime, and heating from the mechanical effects of a giant impact is relatively minor. To estimate the energy released from a giant impact, therefore, we only consider the gravitational potential energy of the mantle \cite{Hussmann2010},
\begin{equation}
    U_{\rm m}=-\frac{16}{3}\pi^{2}G\Bigg[\frac{\bar{\rho}_{\rm m}\bar{\rho}_{\rm c}R^{3}_{\rm c}}{2}\left(R^{2}_{\rm p}-R^{2}_{\rm c}\right) +\frac{\bar{\rho}^{2}_{\rm m}}{5}\left(R^{5}_{\rm p}-R^{5}_{\rm c}\right) -\frac{\bar{\rho}_{\rm m}^{2}R^{3}_{\rm c}}{2}\left(R^{2}_{\rm p}-R^{2}_{\rm c}\right)\Bigg],
\label{eq:U_m}
\end{equation}
where $\bar{\rho}_{\rm c}$ and $R_{\rm c}$ are the average metallic core density and radius respectively. The deposited energy is equal to the difference in the mantle potential energy before and after the impact,
\begin{equation}
    \Delta E =\left|U_{\rm m}\left(M_{\rm n}\right)-U_{\rm m}\left(M_{\rm n}{-}\Delta M\right)\right| = \frac{3GM_{\oplus}^{2}}{5R_{\oplus}} \left(\frac{M_{\rm n}}{M_{\oplus}}\right)^{\frac{7}{4}}\left[1-\left(1-\chi\right)^{\frac{7}{4}}\right]{\rm F_{2}}\left(f_{\rm c}\right),
\label{eq:dE}
\end{equation}
with $\chi$ being the ratio of the mass of the impacting body over the final mass of the planet, and
\begin{equation}
\begin{split}
    {\rm F}_{2}(f_{\rm c})=\left[1+0.3\left(0.325-f_{\rm c}\right)\right]^{-1}\left\{\frac{5}{2}f_{\rm c}\left(1-f_{\rm c}\right)\frac{1-\left(\frac{R_{\rm c}}{R_{\rm n}}\right)^{2}}{1-\left(\frac{R_{\rm c}}{R_{\rm n}}\right)^{3}} \right. & \left.+\left(1-f_{\rm c}\right)^{2}\frac{1-\left(\frac{R_{\rm c}}{R_{\rm n}}\right)^{5}}{\left[1-\left(\frac{R_{\rm c}}{R_{\rm n}}\right)^{3}\right]^{2}} \right. \\
    & \left. -\frac{5}{2}\left(1-f_{\rm c}\right)^{2} \left(\frac{R_{\rm c}}{R_{\rm n}}\right)^{3} \frac{1-\left(\frac{R_{\rm c}}{R_{\rm n}}\right)^{2}}{\left[1-\left(\frac{R_{\rm c}}{R_{\rm n}}\right)^{3}\right]^{2}} \right\}.
\end{split}
\label{eq:F_2}
\end{equation}
Equations~\ref{eq:dE} and \ref{eq:F_2} are obtained from Equation~\ref{eq:U_m} \cite{Sohl2015} after rewriting the layer masses and densities in terms of the total nuclear mass and radius $M_{\rm n}$ and $R_{\rm n}$, metallic core mass fraction $f_{\rm c}$, and metallic core radius $R_{\rm c}$.

We do not consider the energy required to bring the atmosphere into thermodynamic equilibrium with the new surface temperature because this effect is negligible for small atmospheric mass fractions \cite{Biersteker2019}. Accounting for it would further reduce the already limited fraction of impact energy available to drive mass loss, making our estimate conservative. We compare our approach with smoothed particle hydrodynamic simulations of giant impacts \cite{Nakajima2021}: By conducting a series of impact simulations with various target and impactor sizes (though all with $f_{\rm c}$ of 0.3), it has been found that the fraction of impact energy deposited on the target mantle is in the range between ${\sim}0.5$ and ${\sim}0.9$ for the impact angle of 0 to 60 degrees (see Figures 4a and 4c in Ref.\,\cite{Nakajima2021}). To find this fraction with our approach, we also derive the energy gain of the metallic core, by comparing the gravitational binding energy of the metallic core after the giant impact minus that of before as,
\begin{equation}
    \Delta E_{\rm c} = \frac{3GM_{\oplus}^{2}}{5R_{\oplus}} \left(\frac{M_{\rm n}}{M_{\oplus}}\right)^{\frac{7}{4}}\left[1-\left(1-\chi\right)^{\frac{7}{4}}\right]{\rm F_{3}}\left(f_{\rm c}\right),
\label{eq:dEc}
\end{equation}
where
\begin{equation}
    {\rm F}_{3}(f_{\rm c})= \left[1+0.3\left(0.325-f_{\rm c}\right)\right]^{-1} f_{\rm c}^{2}\left(\frac{R_{\rm c}}{R_{\rm n}}\right)^{-1}.
\label{eq:F_3}
\end{equation}
The fraction of impact energy deposited to the mantle is then given by $\Delta E/\left(\Delta E{+}\Delta E_{\rm c}\right){=}1/\left[1{+}{\rm F}_{3}(f_{\rm c})/{\rm F}_{2}(f_{\rm c})\right]$, which depends only on the metallic core mass fraction $f_{\rm c}$. Figure~\ref{fig:mantle}b shows that this fraction is ${\sim}0.84$ for $f_{\rm c}$ of 0.3, which is within the range estimated by Ref.\,\cite{Nakajima2021}.

To find the post-impact temperatures, we create various mantle adiabats with different effective temperatures and determine the energy difference between them. Based on our simulation results, the energy difference, $Q$, is parameterized as follows ($R^{2}{>}0.88$ goodness of fit),
\begin{equation}
    Q=1.75{\times}10^{27}\left(1-f_{\rm c}\right)\left(\frac{M_{\rm n}}{M_{\oplus}}\right)\left(T_{\rm sur}-1610\right)^{\frac{13}{10}},
\label{eq:Q}
\end{equation}
with $T_{\rm sur}$ being the post-impact surface temperature in K. Equation~\ref{eq:Q} is an analytic fit for the energy difference between two mantles of equal mass and composition but different temperature profiles; one having a surface temperature at the rheological transition and the other at temperature $T_{\rm sur}$. This assumes that the surface temperature of the planet before being impacted is set by the rheological transition, which provides a lower-bound estimate for the pre-impact temperature and therefore also for the post-impact temperature inferred from a given deposited energy. Before disk dispersal, gas recycling at the atmosphere-disk interface continually exchanges envelope gas with disk gas and can regulate the thermal evolution of embedded planets \cite{Ormel2015b,Moldenhauer2021,Zhu2021}. For temperate planets, incoming disk gas can carry enough entropy after adiabatic compression in the planetary potential to suppress further cooling \cite{Ormel2015b,Moldenhauer2021,Zhu2021}. For much hotter post-impact envelopes, however, the incoming gas may have insufficient entropy to maintain the shock-heated state, even after adiabatic compression, and recycling can instead contribute to cooling by replacing hot envelope gas with relatively cooler disk gas \cite{Modirrousta2025a}. The effect of recycling therefore depends on the relative thermodynamic properties of the planet and protoplanetary disk, as well as the recycling velocity and flow geometry, and requires dedicated multidimensional modeling to evaluate adequately. After the protoplanetary disk dissipates, gas recycling ceases, but atmospheric thermal blanketing may still slow magma-ocean cooling \cite{Matsui1986,Abe1997}. Even so, a post-disk giant impact can occur millions to tens of millions of years after disk dispersal, leaving time for the planetary surface to cool back toward the rheological transition. After the surface approaches this transition, further cooling becomes inefficient because the magma behaves rheologically like a solid \cite{Solomatov2015}. For simplicity, we therefore treat the rheological-transition temperature as an idealized lower-bound pre-impact state. Combining equations~\ref{eq:dE}{$-$}\ref{eq:Q},
\begin{equation}
    T_{\rm sur}=1610+8500\left(\frac{M_{\rm n}}{M_{\oplus}}\right)^{\frac{15}{26}}\left[1-\left(1-\chi\right)^{\frac{7}{4}} \right]^{\frac{10}{13}}
    \left[\frac{{\rm F}_{2}\left(f_{\rm c}\right)}{1-f_{\rm c}}\right]^{\frac{10}{13}}.
\label{eq:T_sur}
\end{equation}
Equation~\ref{eq:T_sur} gives the surface temperature of a planet of mass $M_{\rm n}$ with metallic core mass fraction $f_{\rm c}$ that experienced a giant impact of mass $\chi M_{\rm n}$. An example with $f_{\rm c} {=} 0.325$ is shown in Figure~\ref{fig:T_sur}. The post-impact surface temperature of a planet can be higher than what Equation~\ref{eq:T_sur} suggests because a giant impact may heat up only a shallow section of the mantle \cite{Nakajima2015}. Equation~\ref{eq:T_sur} therefore serves as an idealized model for mantle heating from giant impacts.

\section*{Data Availability}
The synthetic data used in this study can be generated using the code available on Zenodo \cite{Zenodo2026}. The observational data shown in Figure~\ref{fig:bimodal} and Extended Data Figure~\ref{fig:bimodal_2} are from Ref.~\cite{Fulton2017}.

\section*{Code Availability}
Code to reproduce the results is available on Zenodo \cite{Zenodo2026}.

\section*{Funding Statement}
This work was sponsored by the National Natural Science Foundation of China (grant no. W2632045), the US National Science Foundation EAR-2224727, and the US National Aeronautics and Space Administration under Cooperative Agreement No.\,80NSSC19M0069 issued through the Science Mission Directorate.

\section*{Author Contributions Statement}
D.M.G. and J.K. designed the study and discussed the results. D.M.G. performed calculations and wrote the manuscript.

\section*{Competing Interests Statement}
The authors declare no competing interests.

\clearpage
\renewcommand{\figurename}{Figure}
\setcounter{figure}{0}
\renewcommand{\theHfigure}{main.\arabic{figure}}

\begin{figure}[p]
    \centering
    \includegraphics[width=\linewidth,height=0.74\textheight,keepaspectratio]{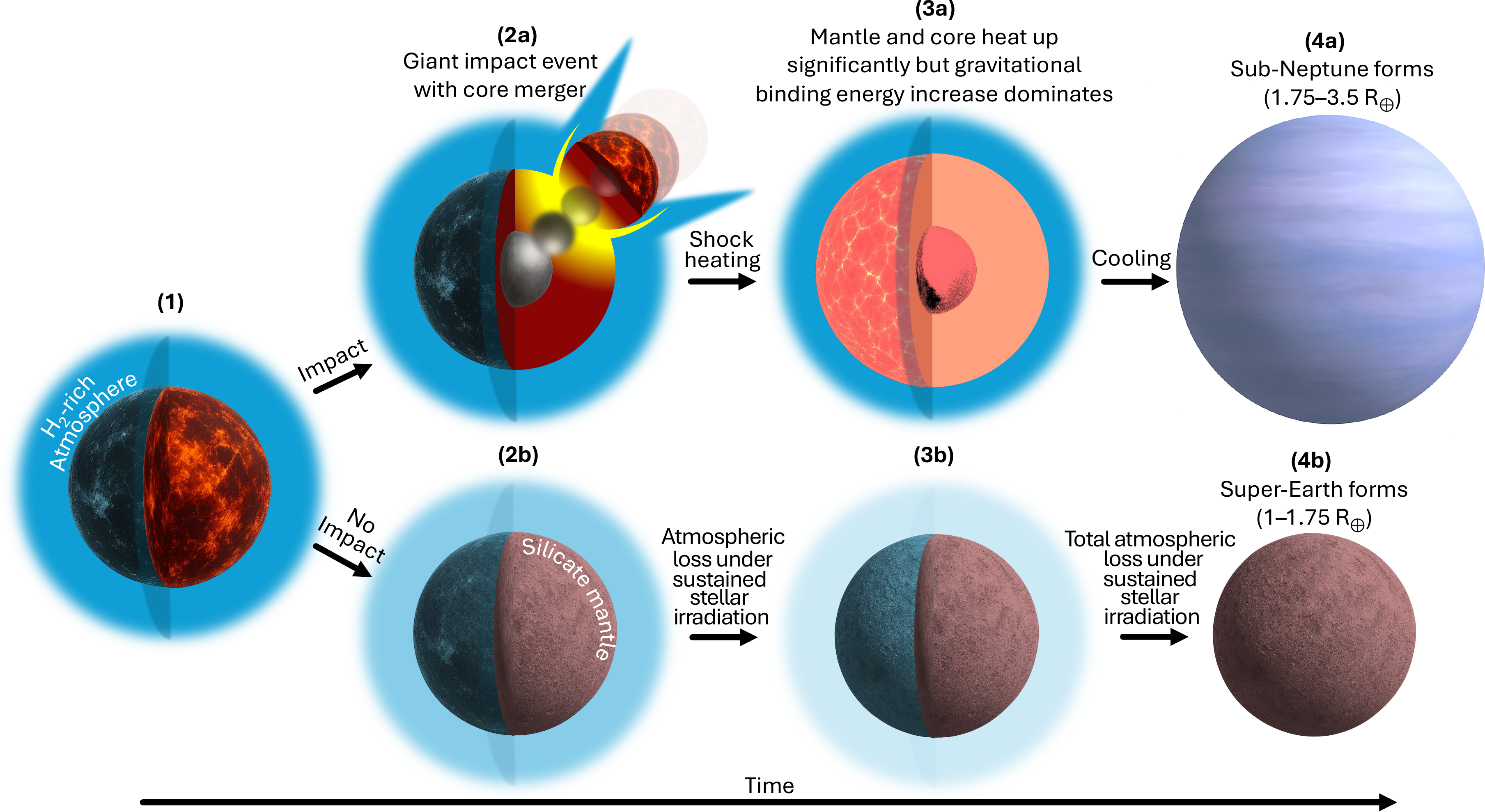}
    \caption{\textbf{Atmospheric evolution with and without a post-disk giant impact.} Panel 1: The common initial state for a rocky planet with a hydrogen-rich atmosphere. Panels 2a–4a: A giant impact deposits energy into the mantle and core, but the increase in gravitational binding energy dominates, allowing the planet to retain its hydrogen-rich atmosphere and evolve into a sub-Neptune ($1.75{-}3.5~R_{\oplus}$). Panels 2b–4b: In the absence of such an impact, the planet is less massive, loses its primordial atmosphere under sustained stellar irradiation, and becomes a super-Earth ($1{-}1.75~R_{\oplus}$).}
    \label{fig:schematic}
\end{figure}
\clearpage

\begin{figure}[p]
    \centering
    \includegraphics[width=\linewidth,height=0.74\textheight,keepaspectratio]{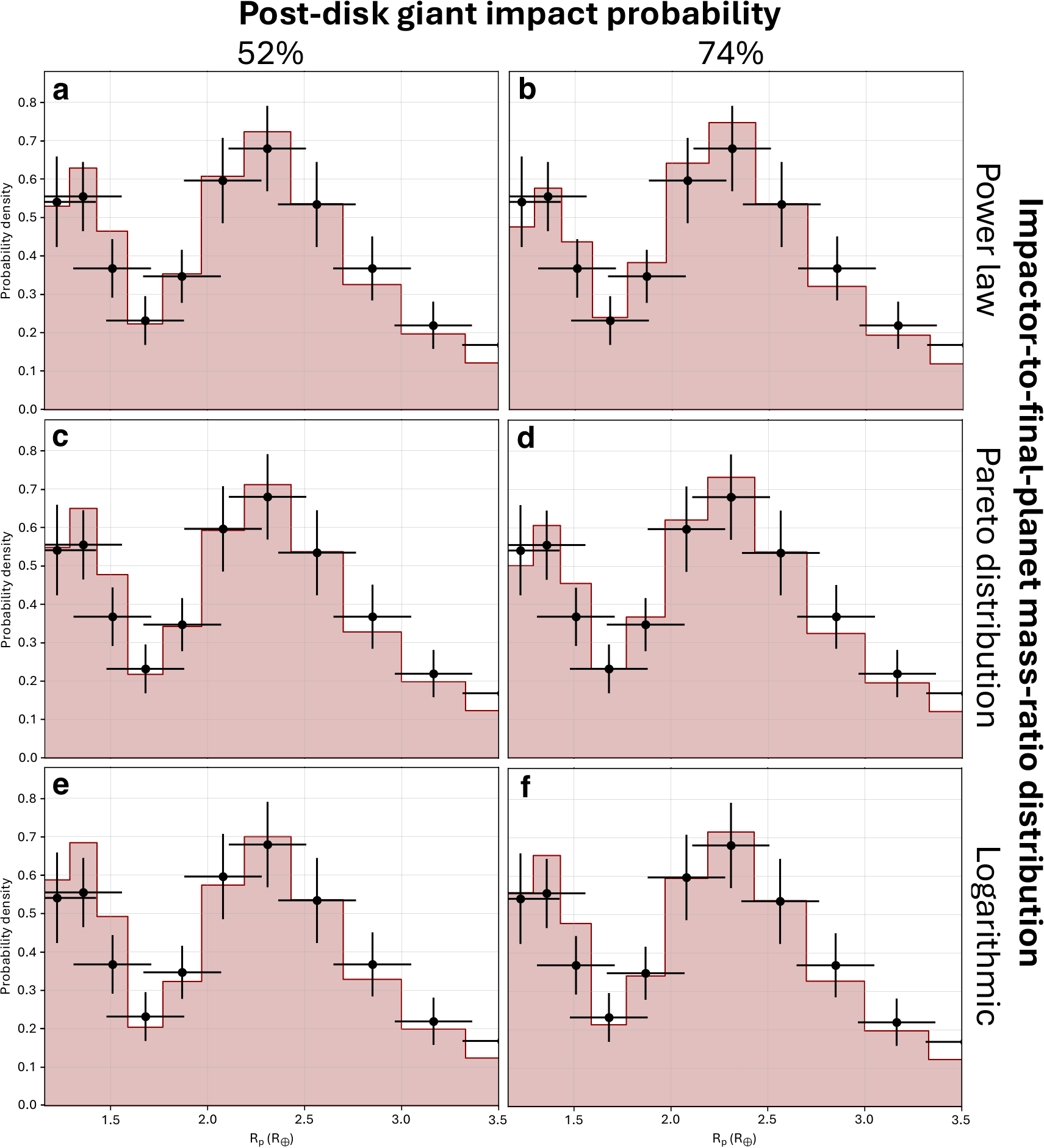}
    \caption{\textbf{Bimodal exoplanet radius distributions under different giant-impact assumptions.} Panels (a,b), (c,d), and (e,f) use the power-law, Pareto, and logarithmic distributions of the impactor-to-final-planet mass ratio defined by Equations~\ref{eq:P_chi}, \ref{eq:P_chi2}, and \ref{eq:P_chi3}, respectively. The left and right columns assume post-disk giant impact probabilities of $52\%$ and $74\%$, respectively, based on N-body simulations \cite{Chance2022,Izidoro2022}. The metallic core mass fraction is uniformly distributed from 10\% to 40\% (based on constraints set by stellar elemental ratios \cite{Unterborn2019} and rocky exoplanet densities \cite{Zeng2016}), and the atmospheric mass fraction scales with nucleus mass according to Equation~\ref{eq:scaling_atm}. The initial exoplanet mass distribution is shown in Figure~\ref{fig:mass_distribution}a. Dark-red step curves and pale-red shading show the probability-density distributions of the final radii of $10^{8}$ synthetic planets. Black circles and vertical bars show the completeness-corrected occurrence in each radius bin and corresponding $1\sigma$ uncertainties from Ref.\,\cite{Fulton2017}, each divided by the area under the binned occurrence distribution. The uncertainties are derived from Poisson counting statistics and scaled using simulated transit surveys. Horizontal bars show a fixed $\pm0.2\,R_{\oplus}$ approximation to the typical radius uncertainty illustrated in Figure 7 of Ref.\,\cite{Fulton2017}.}
    \label{fig:bimodal}
\end{figure}
\clearpage

\begin{figure}[p]
    \centering
    \includegraphics[width=\linewidth,height=0.74\textheight,keepaspectratio]{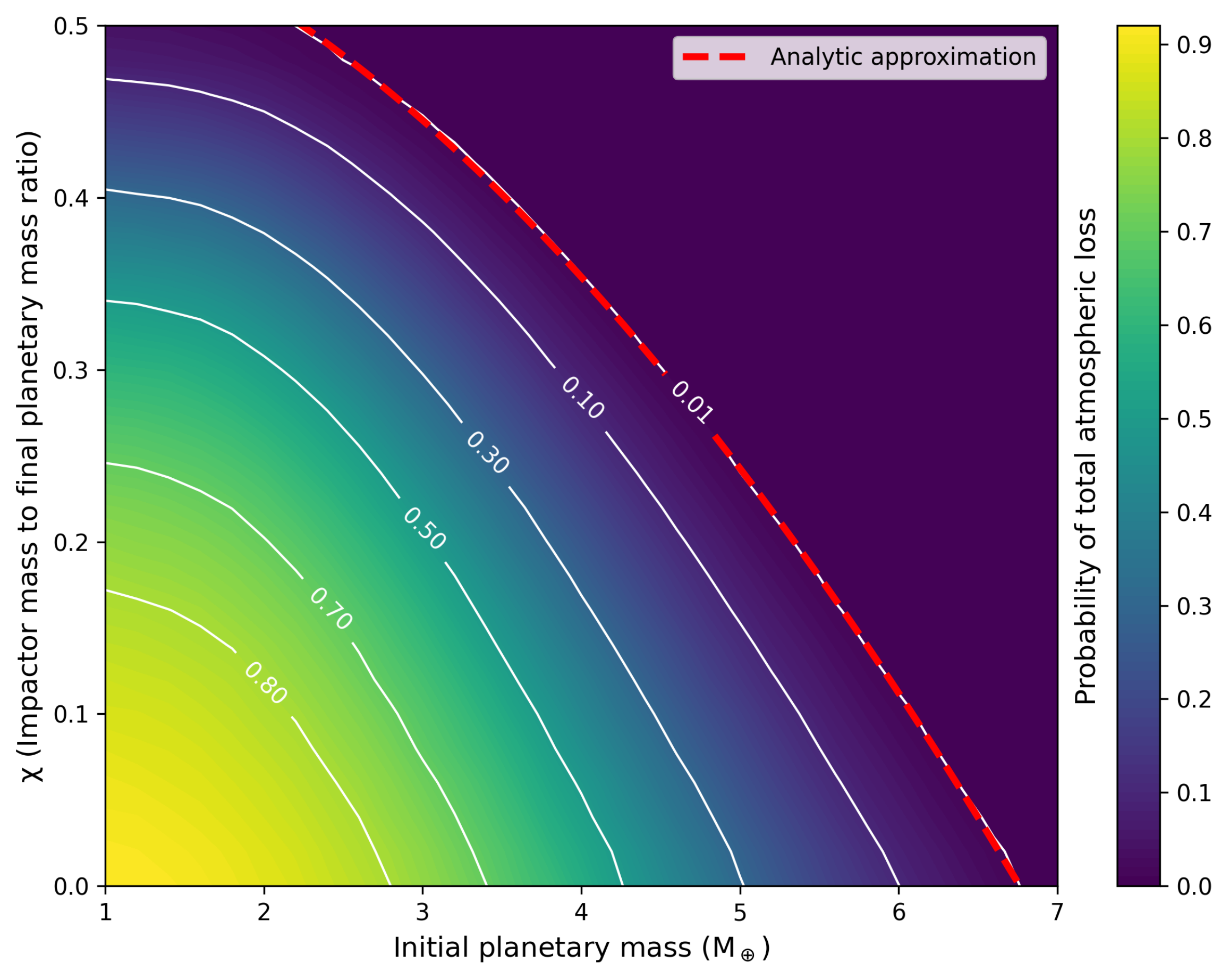}
    \caption{\textbf{Probability of complete atmospheric loss after giant impacts.} The colour scale shows the probability of total atmospheric loss as a function of the impactor-to-final-planet mass ratio, $\chi$, and initial planetary mass, $M_{\rm ini}$. White curves mark probability contours of 0.01, 0.10, 0.30, 0.50, 0.70, and 0.80. This figure was generated using a sample size of $10^{8}$ synthetic planets. The red dashed line shows the analytic approximation for the impactor-to-final-planet mass ratio required for 1\% of all planets of initial mass $M_{\rm ini}$ to experience total atmospheric loss.}
    \label{fig:impact_probability}
\end{figure}
\clearpage

\begin{figure}[p]
    \centering
    \includegraphics[width=\linewidth,height=0.74\textheight,keepaspectratio]{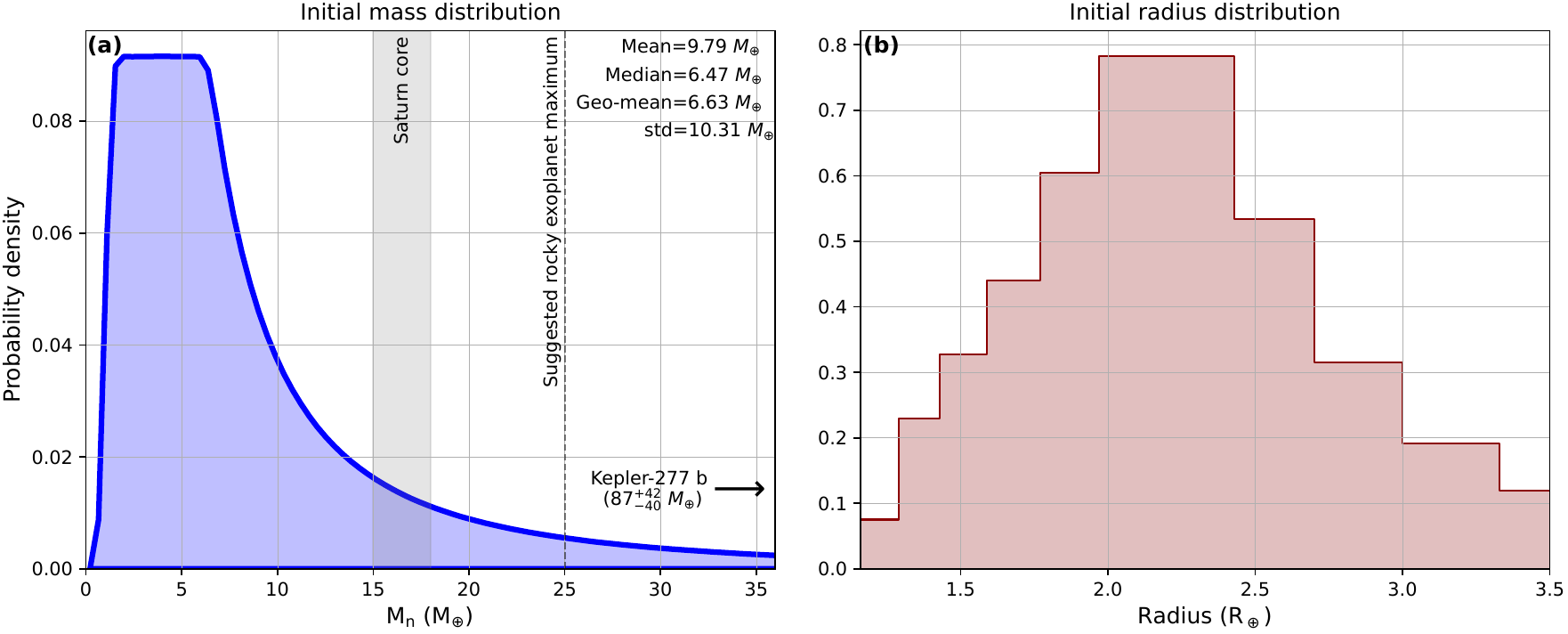}
    \caption{\textls[-5]{\textbf{Adopted initial mass distribution and corresponding radii without atmospheric loss.} (a) The mass distribution of exoplanets adopted in this study. The core of Saturn ($15{-}18~M_{\oplus}$ \protect\cite{Militzer2019,Mankovich2021}) and one of the largest observed rocky exoplanet candidates, Kepler-277~b ($87^{{+}42}_{{-}40}~M_{\oplus}$ \protect\cite{Hadden2014,Xie2014}), are shown for reference. The line at $25~M_{\oplus}$ marks the suggested rocky mass limit proposed by \protect\cite{Otegi2020}. (b) The radius distribution of our synthetic exoplanet population at 5 Gyr, assuming no atmospheric loss occurs.}}
    \label{fig:mass_distribution}
\end{figure}
\clearpage

\begin{figure}[p]
    \centering
    \includegraphics[width=\linewidth,height=0.74\textheight,keepaspectratio]{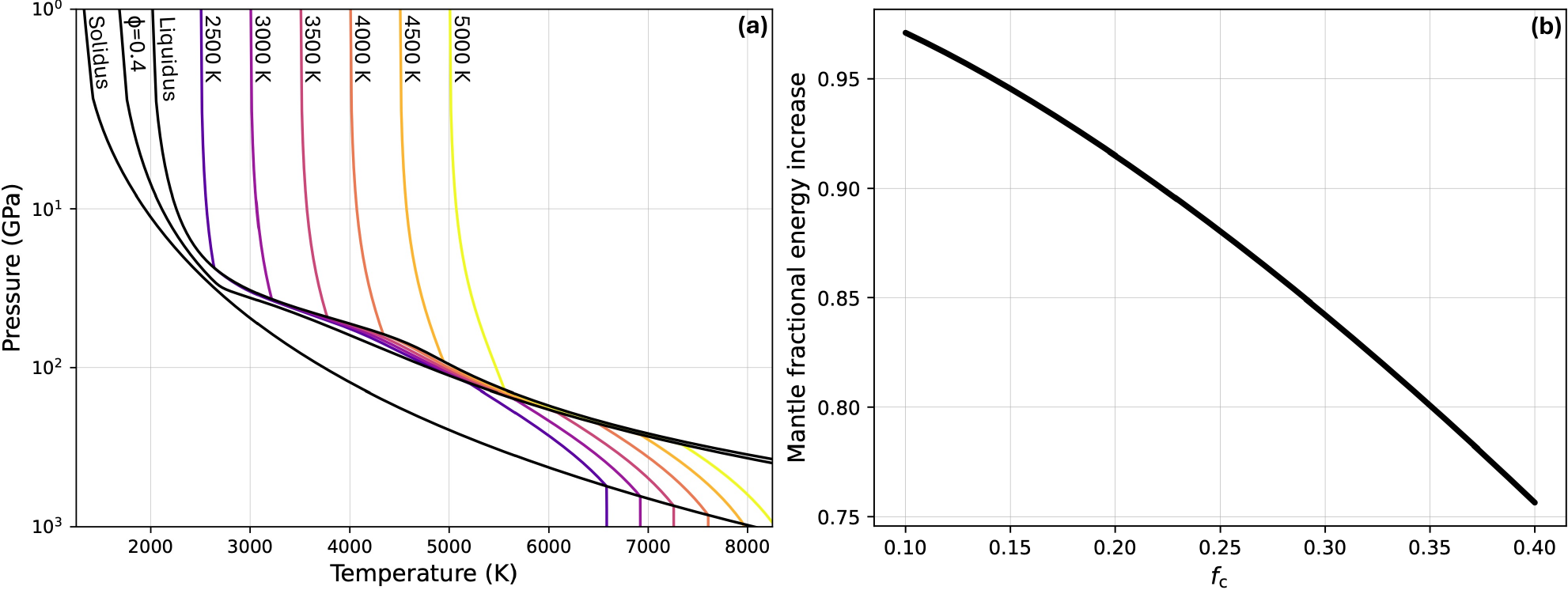}
    \caption{\textbf{Mantle thermodynamics and impact-energy partitioning.} (a) The adiabats for a $10~M_{\oplus}$ planet with a metallic core mass fraction of 0.325 at 2500~K, 3000~K, 3500~K, 4000~K, 4500~K, and 5000~K. The solidus, rheological transition ($\phi{=}0.4$), and liquidus are shown in black. (b) The fraction of impact energy deposited to the mantle, $\Delta E/\left(\Delta E{+}\Delta E_{\rm c}\right)$, as a function of the metallic core mass fraction $f_{\rm c}$.}
    \label{fig:mantle}
\end{figure}
\clearpage

\begin{figure}[p]
    \centering
    \includegraphics[width=\linewidth,height=0.74\textheight,keepaspectratio]{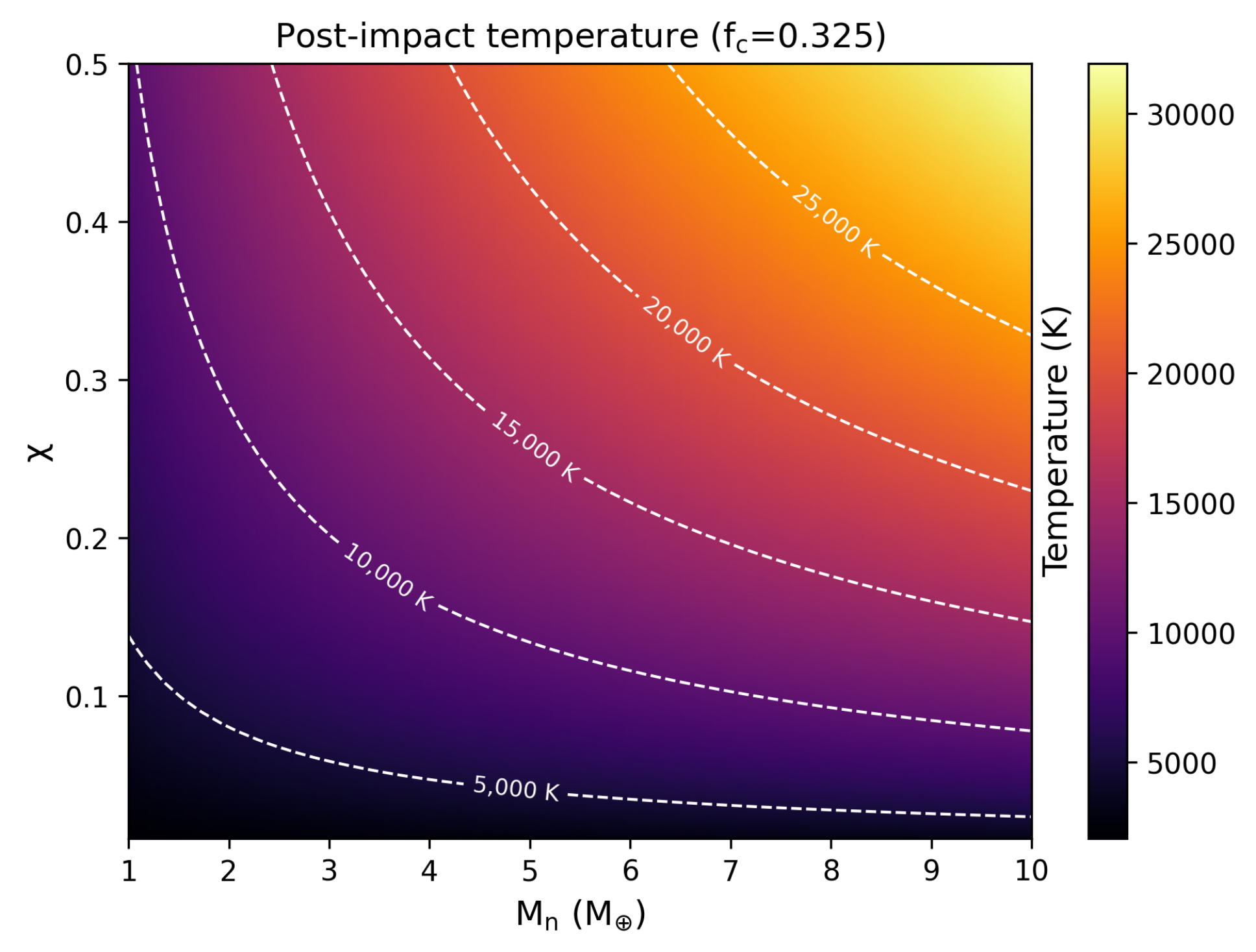}
    \caption{\textbf{Post-impact surface temperatures.} The post-impact surface temperature ($T_{\rm sur}$; Equation~\ref{eq:T_sur}) as a function of the ratio of the impactor’s mass over the final planetary mass ($\chi$) and the planetary nuclear mass ($M_{\rm n}$) for a planet with an Earth-like metallic core mass fraction. White dashed curves mark the labeled temperature contours.}
    \label{fig:T_sur}
\end{figure}
\clearpage

\renewcommand{\figurename}{Extended Figure}
\setcounter{figure}{0}
\renewcommand{\theHfigure}{extended.\arabic{figure}}

\begin{figure}[p]
    \centering
    \includegraphics[width=\linewidth,height=0.74\textheight,keepaspectratio]{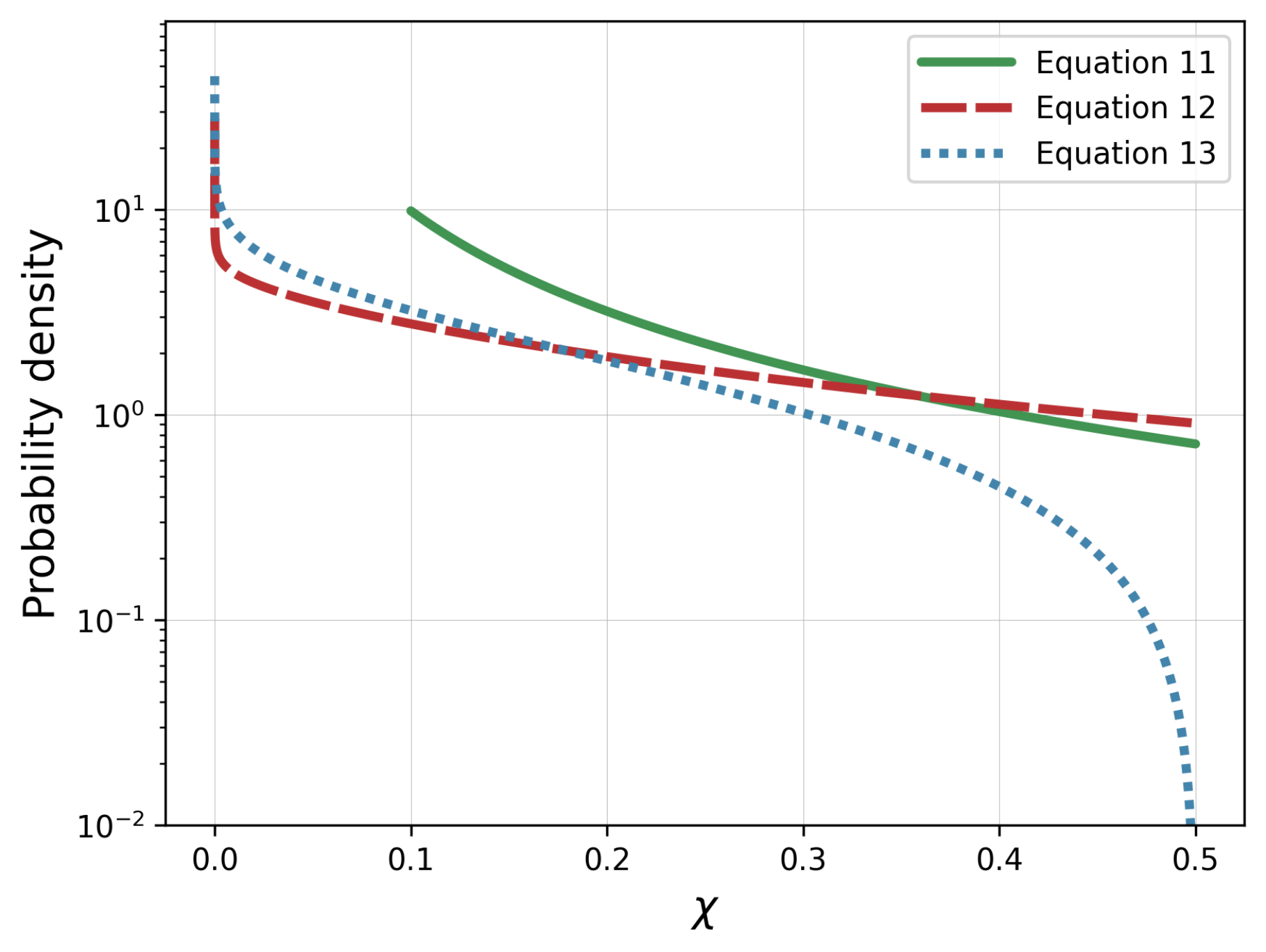}
    \caption{\textbf{Adopted giant-impact size distributions.} The power-law, Pareto, and logarithmic distributions are shown by the green solid, red dashed, and blue dotted curves and are defined by Equations~\ref{eq:P_chi}, \ref{eq:P_chi2}, and \ref{eq:P_chi3}, respectively.}
    \label{fig:chi_distribution}
\end{figure}
\clearpage

\begin{figure}[p]
    \centering
    \includegraphics[width=\linewidth,height=0.74\textheight,keepaspectratio]{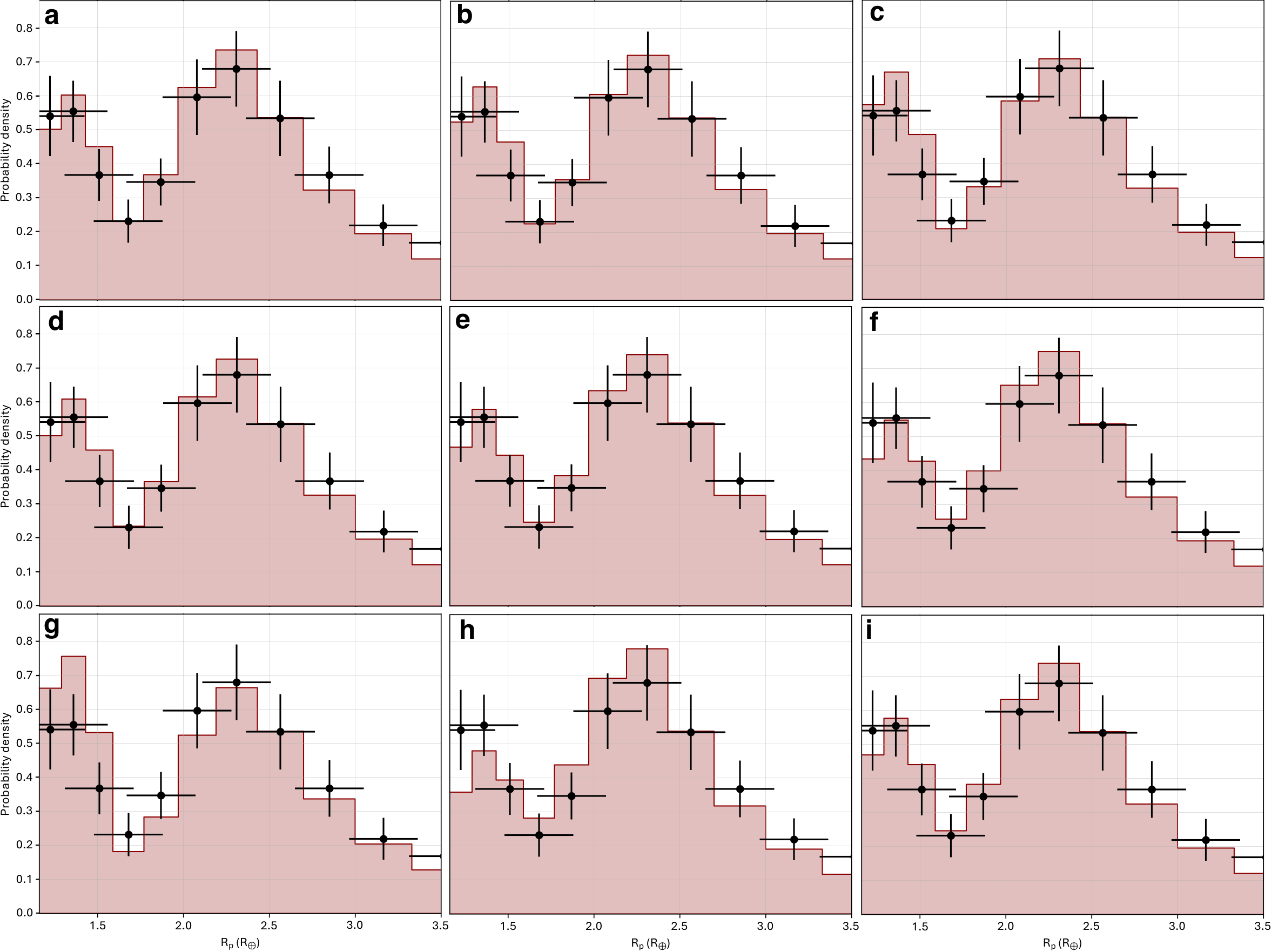}
    \caption{\textbf{Bimodal radius distributions under alternative giant-impact and cooling assumptions.} The points with error bars represent observational data \cite{Fulton2017}. Horizontal uncertainties are approximated from ``typical uncert'' in Figure 7 of Ref.\,\cite{Fulton2017}. Panels (a–c) show results using a $63\%$ impact probability with (a) a power-law, (b) a Pareto, and (c) a logarithmic distribution in $\chi$ (Equations~\ref{eq:P_chi}--\ref{eq:P_chi3}). Panels (d$-$f) adopt a uniform distribution in $\chi$ with post-disk impact probabilities of $52\%$, $63\%$, and $74\%$, respectively. Panels (g$-$i) adopt a uniform distribution in $\chi$. Panels (g) and (h) correspond to post-disk giant impact probabilities of 0\% and 100\%, respectively. Panel (i) corresponds to a 63\% post-disk impact probability, with the magma ocean cooling time $t_{\rm cool}$ set equal to the total planetary age, $t_{\rm age}{=}5~{\rm Gyr}$. Dark-red step curves and pale-red shading show the probability-density distributions of the final radii of $10^{8}$ synthetic planets. Black circles and vertical bars show the completeness-corrected occurrence in each radius bin and corresponding $1\sigma$ uncertainties from Ref.\,\cite{Fulton2017}, each divided by the area under the binned occurrence distribution. The uncertainties are derived from Poisson counting statistics and scaled using simulated transit surveys. Horizontal bars show a fixed $\pm0.2\,R_{\oplus}$ approximation to the typical radius uncertainty illustrated in Figure 7 of Ref.\,\cite{Fulton2017}.}
    \label{fig:bimodal_2}
\end{figure}
\clearpage

\end{document}